\documentclass[10pt,twocolumn,article]{IEEEtran}
\usepackage{times}
\usepackage{amsbsy}
\usepackage{latexsym}
\usepackage{amssymb}
\usepackage{mathtools}
\usepackage{amsmath,amssymb,amsfonts,graphicx,epsfig,epstopdf}
\usepackage{times,amsbsy,latexsym,bm,color}
\usepackage[ansinew]{inputenc}

\title{\huge Study of Multi-Step Knowledge-Aided Iterative Nested MUSIC for Direction Finding \vspace{-0.5em}}
\author{Silvio F. B. Pinto $^1$ and Rodrigo C. de Lamare $^{1,2}$ \\
Center for Telecommunications Studies (CETUC) \\ $^1$ Pontifical
Catholic
University of Rio de Janeiro, RJ, Brazil.\\
$^2$ Department of Electronics, University of York, UK \\
Emails: silviof@cetuc.puc-rio.br, delamare@cetuc.puc-rio.br
\vspace{-0.75em}} 
\begin{document}
\maketitle
\begin{abstract}
In this work, we  propose a  subspace-based algorithm for
direction-of-arrival (DOA) estimation applied to the signals impinging on a two-level nested array, referred to as multi-step
knowledge-aided iterative nested MUSIC method (MS-KAI-Nested-MUSIC), which significantly improves the accuracy of the original Nested-MUSIC. Differently from existing knowledge-aided methods applied to uniform linear arrays (ULAs), which make
use of available known DOAs to improve the estimation of
the covariance matrix of the input data, the proposed Multi-Step
KAI-Nested-MU employs knowledge of the structure of the
augmented sample covariance matrix, which is obtained by exploiting the difference co-array structure
covariance matrix, and its
perturbation terms and the gradual incorporation of prior knowledge,
which is obtained on line. The effectiveness  of the proposed technique can be noticed by
simulations focusing on uncorrelated closely-spaced sources.
\end{abstract}


\section{Introduction}
\label{introduction} Direction-of-arrival (DOA) estimation and
beamforming are two main applications of the sensor array.
Nevertheless, both of them have been mainly restricted to the case
of uniform linear arrays (ULA)
\cite{Vantrees1,locsme,elnashar,manikas,cgbf,okspme,r19,scharf,bar-ness,pados99,
reed98,hua,goldstein,santos,qian,delamarespl07,delamaretsp,xutsa,xu&liu,
kwak,delamareccm,delamareelb,wcccm,delamarecl,delamaresp,delamaretvt,delamaretvt10,delamaretvt2011ST,
delamare_ccmmswf,jidf_echo,jidf,barc,lei09,delamare10,fa10,ccmavf,lei10,jio_ccm,
ccmavf,stap_jio,zhaocheng,zhaocheng2,arh_eusipco,arh_taes,rdrab,dcg,dce,dta_ls,
song,wljio,barc,saalt,mmimo,wence,spa,mbdf,rrmber,bfidd,did,mbthp,wlbd,baplnc}.
The number of sources that can be resolved with a \textit{N} element
ULA using conventional subspace based methods like MUSIC
\cite{Schimdt} is \textit{N-1}. Over the years, the question of
detecting more sources than sensors has been dealt with by different
means. In \cite{Pillai_1,Pillai_2}, the use of minimum redundancy
arrays (MRA) \cite{Moffet} and the construction of  an enlarged
covariance matrix for reaching improved degrees of freedom (DOF)
were not successful in making it positive semidefinite for finite
number of samples. In \cite{Abramovich1,Abramovich2}, an approach to
convert the enlarged matrix into an appropriate positive definite
Toeplitz matrix was proposed. In spite of that, for achieving more
DOF to detect \textit{N-1} sources with \textit{N} sensors, that
approach also depends on MRA, for which there is no closed form
expression for the array geometry. Moreover, such arrays demand hard
designs which are limited to computer simulations or complex
algorithms for locating the sensors
\cite{Johnson,Linebarger},\cite{Chen,Pearson,Ruf}. In \cite{Porat},
the approach using fourth-order cumulants succeeded in increasing
DOF. Yet it is limited to non-Gaussian sources. In \cite{Ma}, by
using the Khatri-Rao (KR) product and the hypothesis of
quasi-stationary sources, one can recognize \textit{2N-1} sources
through a \textit{N} element ULA without needing to calculate
high-order statistics. However, this approach depending on
quasi-stationary sources is not appropriate to stationary sources.
In \cite{Bliss}, the rise of the DOF  resulted from  building  a
virtual array making use of a MIMO radar. Since the creation of that
array relies on active sensing, that method is not suitable for
passive sensing. In \cite{Pal1,Pal2}, by exploring the class of
non-uniform arrays, it was suggested an array structure called
nested array. It is formed by combining two or more ULAs to obtain a
difference co-array, which provides increase of DOF and, therefore,
can resolve more sources than the real number of  physical sensors.
In  a subsequent work \cite{Han}, linear nested arrays were employed
to estimate DOAs  of distributed sources. Additionally, in
\cite{Yang}, it was proposed a robust beamforming for these arrays
based on interference-plus-noise reconstruction and steering vector
estimation. The four last mentioned studies
\cite{Pal1,Pal2,Han,Yang}  focus on scenarios composed of multiple
unclosely spaced sources in order to assess the performances  of
their proposed methods in resolving more sources than the real
number of  physical sensors. For this aim, their  signal models
assume that the sources are uncorrelated. However, the required
vectorization of the initial covariance matrix resulting from the
employment of uncorrelated sources already leads to an equivalent
source signal vector whose  powers of their sources  behave like
fully coherent ones. For this reason, that method, which is based on
a system model assuming uncorrelated sources, makes use of spatial
smoothing.

 In previous works using ULAs \cite{Pinto1,Pinto2,Pinto3,Pinto4}, we developed two ESPRIT-based
 methods known as Two-Step KAI ESPRIT (TS-KAI-ESPRIT), Multi-Step
 KAI-ESPRIT (MS-KAI-ESPRIT) and the Krylov subspace based  Multi-Step
 KAI-Conjugate Gradient (MS-KAI-CG). All of them make use of the refinements of the
 covariance matrix estimates via  steps of reductions
 \cite{Vorobyov1, Vorobyov2} of their undesirable terms. The mentioned methods
 determine the values of scaling factors that reduce the undesirable
 terms causing perturbations in the estimates of the signal and noise
 subspaces in an iterative manner, resulting in better estimates.
 This is carried out by choosing the set of DOA estimates that have
 the highest likelihood of being the set of true DOAs. TS-KAI ESPRIT
 combines this refinement, which has been considered in only two
 steps, with the use of prior knowledge about signals
 \cite{Pinto5,Steinwandt,Stoica1}. Considering a practical scenario,
 the mentioned previous knowledge could be from the signals coming
 from known base stations or from static users in a system. The
 MS-KAI-ESPRIT and MS-KAI-CG, instead of employing prior knowledge about the
 signals, obtain their initial knowledge on line, i.e. by means of
 initial estimates, computed at the first step.  At each iteration of
 their second step, the initial Vandermonde matrix is updated by
 replacing an increasing number of steering vectors of initial
 estimates with their corresponding newer ones. In other words, at
 each iteration, the knowledge obtained on line is updated, allowing
 the correction of the sample covariance matrix estimates, which
 yields more accurate estimates.

 In this work, we  propose a  subspace-based algorithm for
 direction-of-arrival (DOA) estimation applied to the signals impinging on a two-level nested array, referred to as multi-step
 knowledge-aided iterative nested MUSIC method (MS-KAI-Nested-MUSIC), which is inspired by previously reported knowledge-aided
 techniques. Differently from existing knowledge-aided methods applied to uniform linear arrays (ULAs), which make
 use of available known DOAs to improve the estimation of
 the covariance matrix of the input data, the proposed Multi-Step
 KAI-Nested-MU employs knowledge of the structure of the  spatially smoothed
 covariance matrix, which is obtained from processing part of the difference co-array, its
 perturbation terms and the gradual incorporation of prior knowledge,
 which is obtained on line.

  The employment of such ULA-based method like MUSIC in a two-level nested array is justified \cite{Pal1} by the following: its difference coarray, in which is based this method, is a filled longer ULA; the spatially-smoothed covariance matrix resulted from processing signals impinging on a two-level nested array is positive semidefinite for any finite number of snapshots; since its resulting smoothed matrix is  equal to the square of a covariance matrix obtained from the mentioned longer ULA, both covariance matrices share the same  set of eigenvectors and the square of the eigenvalues of the former are equal to the corresponding ones of the later.

   This paper is organized as follows. Section \ref{sysmodel} briefly describes the Nested-MUSIC and the necessary background for understanding the proposed technique. Section \ref{propos_MS_KAI_nest_MU} presents the proposed MS-KAI-Nested-MUSIC algorithm. Section \ref{computational_analysis}, illustrates and discusses the computational complexity of the proposed algorithm. In Section \ref{simulations}, we present the simulation results whereas the conclusions are drawn in Section \ref{conclusions}.

 \textit{Notation}: the superscript \textit{H} denote the Hermitian transposition,  $\mathbb E[\cdot]$ expresses the expectation operator  and $\bm I_M$ stands for the $M\times M$  identity matrix.

\section{System Models and Background}
\label{sysmodel}

\subsection{The Nested MUSIC system model}
\label{nested_MU_sys_mod}
\begin{figure}[!h]
    \centering 
    \includegraphics[width=8.3cm,height=2.2cm]{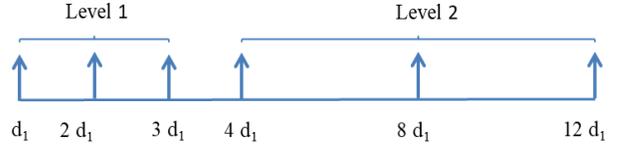} 
    \vspace{-1.0em}\caption{A two level nested array with 3 sensors at each level.}
    \label{figura:nest_2levels_arrangement}
\end{figure}

Let us consider a two-level nested passive array composed of $\mathit{M}$ sensors, which is a concatenation of two ULAs. The inner ULA has $\mathit{M_{1}}$ sensors with intersensor spacing $\mathit{d}_{1}$ and the outer
has $\mathit{M_{2}}$ sensors with intersensor spacing $\mathit{d}_{2}=\left(\mathit{M_{1}}+1 \right)\mathit{d}_{1} $. Specifically speaking, it is a linear array with sensors positions obtained by the union of the sets $I_{nner}= \left\lbrace md_{1}\mid\: m=1, 2, \ldots, M_{1}\right\rbrace $ and $O_{uter}= \left\lbrace n\left\lbrace M_{1}+1 \right\rbrace d_{1}\mid\: n=1, 2, \ldots, M_{2}\right\rbrace $. Fig.     \ref{figura:nest_2levels_arrangement} illustrates a two-level nested array.
\vspace{-1.0em}

  Assuming $\mathit{P}$ uncorrelated narrowband signals from far-field sources at directions $\left\lbrace\theta_{p}, p=1,2,\ldots,P \right\rbrace $ impinging on this array, the  $i$th data snapshot of the $M$-dimensional array output vector
can be modeled as
\begin{equation}
\label{model_nest}
\bm y(i)=\bm {F\,s}(i)+\bm n(i),\qquad i=1,2,\ldots,N,
\end{equation}
where $\bm y(i)=[y_{1}(i),y_{2}(i),\ldots,y_{M}(i)]^T$ is the received signal vector at the snapshot $\mathit{i}$, $\bm s(i)=[s_{1}(i),s_{2}(i),\ldots,s_{P}(i)]^T$ is the source signal vector and $s_{p}(i)\sim N^{C}\left(0, \sigma_{k}^{2} \right) $. Additionally, we assume that $\bm n(i)=[n_{1}(i),n_{2}(i),\ldots,n_{M}(i)]^T$ is the white Gaussian noise vector with power $\sigma_{n}^{2}$ and that its components and the source vector ones are uncorrelated to each other. We also consider that
 $\bm {f}(\theta_p)=\left\lbrace e^{-j2\pi\frac{d_{1}}{\lambda_{c}}r_{n}\sin\theta_{p}}\mid n=1,2,\ldots,M\right\rbrace $ denotes the steering vector of the \textit{pth} signal, where $\lambda_{c}$ stands for the carrier wavelength and
 \begin{align}
 \left\lbrace r_{n} \mid n=1,2,\ldots,M\right\rbrace =&\left\lbrace 0,1,\ldots,M_{1}-1, M_{1},\right.\nonumber\\
 &\left.2\left(M_{1}+1\right)-1,\ldots,\right.\nonumber\\
 &\left. M_{2}\left(M_{1}+1 \right)-1\right\rbrace
 \end{align}
is a vector that contains the location of the sensors. Next, the array manifold containing the steering vectors of the signals can be formed as
\begin{equation}
\label{array_manif_nest}
\bm {F}(\Theta)=[\bm
{f}(\theta_{1}),\bm
{f}(\theta_{2}),\ldots,\bm{f}(\theta_{P})]
\end{equation}
By averaging the \textit{N} collected snapshots through the time, we can express the sample covariance matrix as

\begin{eqnarray}
\label{covsample_nest1}
\bm {\hat{R}_{1}}=\frac{1}{N} \sum\limits^{N}_{i=1}\bm{y}\left( i\right) \bm{y} ^H\left( i\right) \approx \mathbb E\left[\bm{y}\left( i\right) \bm{y} ^H\left( i\right)
\right]\nonumber\\
=\bm {F}\,\bm R_{s}\bm{F} ^H+
\sigma_n^2\bm I,
\end{eqnarray}
where $\bm{R}_{s}= \mathit{diag}\left\lbrace \sigma_{1}^{2},\sigma_{2}^{2},\ldots,\sigma_{P}^{2}\right\rbrace $

By vectorizing $ \bm {\hat{R}_{1}} $ \eqref{covsample_nest1}, we can obtain a long vector, in which some elements appear more than once. By removing these repeated rows  and sorting them so that the \textit{ith} row corresponds to the sensor located at $\left(-\bar{M}+i \right)d_{1} $,  where   $\bar{M} =\left(M^{2}/4 +M/2 \right)$, we can obtain a new vector
\begin{eqnarray}
\bm {z}=\bm {G}\:\bm{p} +
\sigma_n^2\bm{e},
\label{big_vec_nest}
\end{eqnarray}
where
\begin{equation}
\label{array_manif_nest2}
\bm {G}(\Theta)=[\bm
{g}(\theta_{1}),\bm
{g}(\theta_{2}),\ldots,\bm{g}(\theta_{P})],
\end{equation}
in which
\begin{align}
\bm{g}(\theta_{p})=&\left[e^{-j2\pi\frac{d_{1}}{\lambda_{c}}\left(-\bar{M}+1 \right) \sin\theta_{p}}, e^{-j2\pi\frac{d_{1}}{\lambda_{c}}\left(-\bar{M}+2 \right) \sin\theta_{p}},\right.\nonumber\\
&\left.\ldots, e^{-j2\pi\frac{d_{1}}{\lambda_{c}}\left(\bar{M}-2 \right) \sin\theta_{p}}, e^{-j2\pi\frac{d_{1}}{\lambda_{c}}\left(\bar{M}-1 \right) \sin\theta_{p}} \right]^{T},
\end{align}
\begin{equation}
\label{equiv_source_signal}
\bm{p}=\left[\sigma_{1}^{2}, \sigma_{2}^{2},\ldots,\sigma_{P}^{2} \right]^{T},
\end{equation}
and
\begin{equation}
\bm{e}\in\mathfrak{R}^{\left( 2\bar{M}-1\right)\times 1 }
\end{equation}
is a vector of all zeros, except for a $\mathit{1}$ at the center position.

By comparing \eqref{big_vec_nest}  with \eqref{model_nest}, we can notice that $ \bm {z} $ in \eqref{big_vec_nest} behaves like the signal received by a longer difference coarray, whose sensors locations can be determined by the distinct values in the set $\left\lbrace r_{i}-r_{j}\mid\:1\leq i, j\leq M \right\rbrace $. The equivalent source signal vector $ \bm{p} $ \eqref{equiv_source_signal} consists  of powers $ \sigma_{p}^{2} $ of the actual sources and thus they
behaves like fully coherent sources\cite{Pal1}. This, combined with the fact that the  difference coarray is a filled ULA, motivates to apply spatial smoothing to $\bm {z}$ \eqref{big_vec_nest} to obtain a full rank covariance matrix $ \widetilde{R} $, as follows:

\begin{eqnarray}
\label{covsample_nest_smooth}
\bm{\widetilde{R}} =&\frac{1}{M^{2}/4+M/2} \sum\limits^{M^{2}/4+M/2}_{i=1}\bm{z}_{i}\bm{z}_{i} ^H \nonumber\\
=&\frac{1}{M^{2}/4+M/2}\left( \bm {G}_{1}\bm R_{s}\bm{G}_{1} ^H+
\sigma_n^2\bm I\right)^{2},
\end{eqnarray}
where $\bm{z}_{i} $ corresponds to the $\left( M^{2}/4+M/2-i+1\right)$ \textit{th} to
$ \left( \left( M^{2}-2\right)/2+M -i+1\right) $ \textit{th} rows of $\bm{z}$ and $\bm {G}_{1} $  is a manifold array composed of the last $\bar{M}$ rows of $\bm {G}$.

It can be shown \cite{Pal1} that the smoothed covariance matrix $\bm {\widetilde{R}}$ \eqref{covsample_nest_smooth} can be expressed as $\bm {\widetilde{R}}=\bm {\hat{R}}^{2}$, where $ \bm {\hat{R}} $ has the same form as the covariance received by a longer ULA composed of $M^{2}/4+M/2 $  sensors. Since $ \bm {\hat{R}} $ and $\bm {\widetilde{R}}$ share the same set of eigenvectors and the eigenvalues of $ \bm {\hat{R}}$ are the square roots of $\bm {\widetilde{R}}$, by eigendecomposition of  $\bm {\widetilde{R}}$, we can found the eigenvectors corresponding to the smallest $ M^{2}/4+M/2-P $ eigenvalues of $ \bm {\hat{R}} $.
Due to the previously mentioned reasons and also for being PSD by construction, which results from the sum of vector outer products, the spatially smoothed matrix $\bm {\widetilde{R}}$ can be used as the basis for our proposed MS-KAI-Nested-MUSIC algorithm.

\subsection{Background - ULA model and MUSIC algorithm}
\label{background}
Let us assume that \textit{P}  narrowband signals from
far-field sources impinge on a uniform linear array (ULA) of $M_{1}> \textit{P}$ sensor elements from  directions ${\bm
    \theta}=[\theta_{1},\theta_{2},\ldots, \theta_P]^T$. We also
consider that the sensors are spaced from each other by a distance $
d_{1}\leq\frac{\lambda_{c}}{2}$, where $\lambda_{c}$ is the signal wavelength,
and that without loss of generality, we have
\text{${\frac{-\pi}{2}\leq\theta_{1}\leq\theta_{2}\ldots
        \leq\theta_P\leq \frac{\pi}{2}}$}.

The $i$th data snapshot of the $M_{1}$-dimensional array output vector
can be modeled as
\begin{equation}
\bm {x} \left( i\right) =\bm {A\,s}\left( i\right) +\bm{n} \left( i\right) ,\qquad i=1,2,\ldots,N,
\label{model}
\end{equation}
where $\bm s(i)=[s_{1}(i),\ldots,s_{P}(i)]^T
\in\mathbb{C}^{\mathit{P\times1}}$ represents the zero-mean source
data vector, $\bm n(i) \in\mathbb{C}^{\mathit{M_{1} \times 1}}$ is the
vector of white circular complex Gaussian noise with zero mean and
variance $\sigma_n^2$, and $N$ denotes the number of available
snapshots. The Vandermonde matrix $\bm A \left( \Theta\right) =[\bm
a(\theta_{1}),\ldots,\bm a(\theta_{P})] \in\mathbb
{C}^{\mathit{M_{1}\times P}}$, known as the array manifold, contains the
array steering vectors $\bm a(\theta_j)$ corresponding to the $n$th
source, which can be expressed as
\begin{equation}
\bm a(\theta_n)=[1,e^{-j2\pi\frac{d}{\lambda_{c}}
    \sin\theta_n},\ldots,e^{-j2\pi(M_{1}-1)\frac{d}{\lambda_{c}}\sin\theta_n}]^T,
\label{steer}
\end{equation}
where $n=1,\ldots, P$. Using the fact that $\bm {s}\left( i\right) $ and $\bm {n}\left( i\right) $
are modeled as uncorrelated linearly independent variables, the
$M_{1}\times M_{1}$ signal covariance matrix is calculated by

\begin{equation}
\bm {R } =\mathbb {E}\left[\bm {x}\left( i\right) \bm {x}^H\left( i\right)
\right]=\bm {A}\,\bm {R}_{ss}\bm {A}^H+
\sigma_n^2\bm {I}_M,
\label{covariance}
\end{equation}
where $\bm {R}_{ss}=$ $\mathbb {E}\left[\bm {s}\left( i\right) \bm {s}^H\left( i\right)\right] =$ diag $\left\lbrace \sigma_{1}^{2}, \sigma_{2}^{2}, \ldots,\sigma_{P}^{2}\right\rbrace $.
 Since the true signal covariance matrix is unknown,
it must be estimated and a widely-adopted approach is the sample
average formula given by
\begin{equation}
\bm {\hat{R}}=\frac{1}{N} \sum\limits^{N}_{i=1}\bm x(i)\bm x^H(i),
\label{covsample}
\end{equation}
whose estimation accuracy is dependent on $N$.

From \cite{Schimdt}, it is known that $\bm{R}$  \eqref{covariance} has $M_{1}$ eigenvalues $\left[\lambda_{1},\lambda_{2}, \ldots, \lambda_{P} \right]$ and $ M_{1} $ associated eigenvectors forming a subspace $\bm{\Phi}=\left[ \bm{\phi}_{1},\bm{\phi}_{2},\ldots, \bm{\phi}_{M_{1}}\right] $. By sorting the $ M_{1} $ eigenvalues from the smallest to the largest, the subspace
$\bm{\Phi}$ can be decomposed into two subspaces $\bm{\Phi}= \left[ \bm{\Phi}_{S}\:\bm{\Phi}_{N}\right] $, where $ \bm{\Phi}_{S} $ is the $ \left[ M_{1}\times P\right] $  signal subspace composed of the eigenvectors associated with the impinging signals  and $ \bm{\Phi}_{N}$ is the  $ \left[ M_{1}\times\left(M_{1} - P \right) \right] $  noise subspace, composed of the eigenvectors associated with the noise.
Due to the orthogonality  between  the noise subspace and the array steering vector at the angles of arrival $\left(\theta_{1}, \theta_{2},\ldots,\theta_{P}, \right) $, the matrix  product $ \bm {a}^{H}\left(\theta \right)\bm{\Phi}_{N}\bm{\Phi}_{N}^{H} \bm {a}^{H}\left(\theta \right) $ tends to zero. The reciprocal of this matrix product $ \left( \bm {a}^{H}\left(\theta \right)\bm{\Phi}_{N}\bm{\Phi}_{N}^{H} \bm {a}^{H}\left(\theta \right)\right)^{-1}  $ called MUSIC pseudospectrum creates sharp peaks at the angles of arrival. By plotting it in the range $ \left[-\pi/2\quad \pi/2 \right] $, it is possible to determine the peaks and its corresponding angles of arriving by a peak search.

\section{ The proposed MS-KAI-Nested-MUSIC algorithm}
\label{propos_MS_KAI_nest_MU}
The idea behind the MS-KAI-Nested-MUSIC algorithm is to expand
  the estimated spatially smoothed covariance matrix $\bm {\widetilde{R}}$ \eqref{covsample_nest_smooth} as if it were generated by the \textit{ith} data snapshots of  $L= M^{2}/4+M/2$-dimensional array output vectors, where, as mentioned in \ref{nested_MU_sys_mod}, $ M $ is the number of the physical sensors of the nested array. That is to say that we can employ the estimated spatially smoothed covariance matrix $\bm {\widetilde{R}}$ as if it were the estimate provided by the sample  average formula.
  Therefore, after making $\bm {\widetilde{R}}$ \eqref{covsample_nest_smooth} equal to $\bm {\hat{R}}$ \eqref{covsample} , we can start by   expanding the former \eqref{covsample_nest_smooth}
using \eqref{model} as follows:

\begin{eqnarray}
\bm {\widetilde{R}}=\frac{1}{N} \sum\limits^{N}_{i=1}(\bm A\,s(i)+\bm
n(i))\:(\bm A\,s(i)+\bm n(i))^H \nonumber\\= \bm
A\left\lbrace\frac{1}{N} \sum\limits^{N}_{i=1}\bm s(i)\bm
s^H(i)\right\rbrace\bm A^H+\:\frac{1}{N} \sum\limits^{N}_{i=1}\bm
n(i)\bm n^H(i)\;+\nonumber\\\underbrace{\bm A\left\lbrace\frac{1}{N}
    \sum\limits^{N}_{i=1}\bm s(i)\bm n^H(i)\right\rbrace\:
    +\:\left\lbrace\frac{1}{N} \sum\limits^{N}_{i=1}\bm n(i)\bm
    s^H(i)\right\rbrace\bm{A}^{H}}_{\text{"undesirable by-products"}}
\label{expandedcovsample}
\end{eqnarray}
The first two terms of \text{$\bm {\widetilde{R}}$} in
\eqref{expandedcovsample} can be considered as estimates of the two
summands of \text{$\bm R$} given in  \eqref{covariance}, which
represent the signal and the noise components, respectively.  The
last two terms in \eqref{expandedcovsample} are undesirable
by-products, which  can be seen as estimates for the correlation
between the signal and the noise vectors. The system model under
study is based on noise vectors which are zero-mean  and also
independent of the signal vectors. Thus, the signal and noise
components are uncorrelated to each other. As a consequence, for a
large enough number of samples $N$, the last two  terms pointed out
in \eqref{expandedcovsample} tend to zero. Nevertheless, in practice
the number of available samples can be limited. In such situations,
the last two terms in \eqref{expandedcovsample} may have significant
values, which causes the deviation of the estimates of the signal
and the noise subspaces from the true signal and noise ones. The key
point of the proposed MS-KAI-Nested-MUSIC algorithm is to modify the smoothed covariance matrix estimate at each iteration by gradually
incorporating the knowledge provided by the updated Vandermonde
matrices which progressively incorporate the newer estimates from
the preceding iteration. Based on these updated Vandermonde
matrices, refined estimates of the projection matrices of the signal
and noise subspaces are calculated. These estimates of projection
matrices associated with the initial smoothed covariance matrix estimate and the reliability factor employed to reduce its
by-products allow to choose the set of estimates that has the
minimum value of the SMLOF, i.e., the highest likelihood of being
the set of the true DOAs. The modified smoothed covariance matrix estimate is computed
by deriving a scaled version of the undesirable terms from $\bm
{\widetilde{R}}, $ which are pointed out in \eqref{expandedcovsample}.

The steps of the proposed algorithm are listed in Table
\ref{tab_MSKAI_Nes_MU}. The algorithm starts by computing the
spatially smoothed covariance matrix estimate \eqref{covsample_nest_smooth}. Next, based on it, the DOAs are
estimated using the original MUSIC \cite{Schimdt} algorithm, as briefly described in \ref{background}. The superscript
$(\cdot)^{(1)}$ refers to the estimation task performed in the first
step. Now, a process composed of $n=1:I$ iterations starts by
forming the Vandermonde matrix using the DOA estimates. Then, the
amplitudes of the sources are estimated such that the square norm of
the differences between the observation vector and the vector
containing estimates and the available known DOAs is minimized. This
problem can be formulated as
\begin{eqnarray}
    \hat{\bm{s}}(i)=\arg\min_{\substack{\bm
    s}}\parallel\bm{x}(i)-\hat{\bm{A}}\mathbf{s}\parallel^2_2.
    \label{minimization1}
\end{eqnarray}
The minimization of \eqref{minimization1} is achieved using the
least squares technique and the solution is described by
\begin{equation}
\hat{\bm{s}}(i)=(\mathbf{\hat{A}}^{H}\:\mathbf{\hat{A}})^{-1}\:\mathbf{\hat{A}}\:\bm{x}(i)
\label{minimization2}
\end{equation}
The noise component is then estimated as the difference between the
estimated signal and the observations made by the array, as given by
\begin{eqnarray}
 \hat{\bm n}(i)=\bm x(i)\:-\: \hat{\bm A}\:\hat{\bm s}(i).
\label{noise_component}
\end{eqnarray}
After estimating  the signal and noise vectors, the third term in
\eqref{expandedcovsample} can be computed as
\begin{align}
\bm{V}&\triangleq \hat{\bm{A}}\left\lbrace\frac{1}{N}
\sum\limits^{N}_{i=1}\bm \hat{\mathbf{s}}(i)\bm
\hat{\mathbf{n}}^H(i)\right\rbrace\nonumber\\&=\hat{\bm{A}}\left\lbrace\frac{1}{N}
\sum\limits^{N}_{i=1}(\mathbf{\hat{A}}^{H}\:\mathbf{\hat{A}})^{-1}\mathbf{\hat{A}}^{H}\bm{x}(i)\right.\nonumber\\&\left.\times(\bm{x}^{H}(i)-\bm{x}^{H}(i)\hat{\mathbf{A}}(\hat{\mathbf{A}}^{H}\hat{\mathbf{A}})^{-1}\:\hat{\mathbf{A}}^{H})\right\rbrace\nonumber\\&=\mathbf{\hat{Q}}_{A}\left\lbrace\frac{1}{N}
\sum\limits^{N}_{i=1}
\bm{x}(i)\bm{x}^H(i)\:\left(\mathbf{I}_{M}\:-\:\hat{\mathbf{Q}}_{A}\right)
\right\rbrace\nonumber\\&=\mathbf{\hat{Q}}_{A}\:\mathbf{\widetilde{R}}\:\mathbf{\hat{Q}}_{A}^{\perp},
\label{terms_deducted}
\end{align}
where
\begin{equation}
\mathbf{\hat{Q}}_{A}\triangleq \mathbf{\hat{A}}\:(\mathbf{\hat{A}}^{H}\:\mathbf{\hat{A}})^{-1}\:\mathbf{\hat{A}}^{H}
\end{equation}
is an estimate of the projection matrix of the signal subspace, and
\begin{equation}
\mathbf{\hat{Q}}_{A}^{\perp}\triangleq\mathbf{I}_{M}\:-\:\mathbf{\hat{Q}}_{A}
\end{equation}
is an estimate of the projection matrix of the noise subspace.

Next, as part of the process of $n=1:I $ iterations, the modified
data covariance matrix $\mathbf{\widetilde{R}}^{(n+1)}$ is calculated by
computing a scaled version of the estimated terms from the initial
smoothed covariance matrix as given
\begin{equation}
\label{modified_data_covariance}
\mathbf{\widetilde{R}}^{(n+1)} = \mathbf{\widetilde{R}}\:-\:\mathrm{\mu}\:(\mathbf{V}^{(n)}\:+\:\mathbf{V}^{(n)H}),
\end{equation}
where the superscript $(\cdot)^{(n)}$ refers to the $n^{th} $
iteration performed. The scaling or reliability factor \text{$\mu $}
increases from 0 to 1 incrementally, resulting in modified smoothed covariance matrix estimates. Each of them gives origin to new estimated DOAs
also denoted by the superscript  $(\cdot)^{(n+1)}$ by using the
MUSIC algorithm, as briefly described in \ref{background}.

In this work, the rank \textit{P} is assumed to be
known, which is an assumption frequently found in the literature.
Alternatively, the rank \textit{P} could be estimated by model-order
selection schemes \cite{Rappaport} such as Akaike's Information Theoretic Criterion
(AIC) \cite{Schell} and the Minimum Descriptive Length (MDL)
Criterion  \cite{Rissanen}.

Then, a new Vandermonde matrix $\mathbf{\hat{B}}^{(n+1)}$ is formed
by the steering vectors of those newer estimated DOAs. By using this
new matrix, it is possible to compute the newer estimates of the
projection matrices of the signal \text{$
\mathbf{\hat{Q}}_{B}^{(n+1)} $} and the noise \text{$
\mathbf{\hat{Q}}_{B}^{(n+1)\perp} $} subspaces.

Afterwards, employing the newer estimates of the projection matrices, the
initial smoothed covariance matrix estimate, $\bm {\hat{R}}$,  the number of
its corresponding sensors and the number of sources, the stochastic maximum likelihood objective
function $\mathit{U^{(n+1)}(\mu)}$ \cite{Stoica2} is computed for
each value of \text{$\mu $} at the $n^{th}$ iteration, as follows:
\begin{equation}
\mathit{U^{(n+1)}(\mu)}=\mathrm{ln\:det}\left(\cdot\right),
\label{SML_objective_function}
\end{equation}
where
\begin{equation}
\left(\cdot\right) =\left(\mathbf{\hat{Q}}_{B}^{(n+1)}\:\mathbf{\widetilde{R}}\:\mathbf{\hat{Q}}_{B}^{(n+1)}+\dfrac{{\rm Trace}\{\mathbf{\hat{Q}}_{B}^{\perp\:(n+1)}\:\mathbf{\widetilde{R}}\}} {\mathrm{L-P}}\:\mathbf{\hat{Q}}_{B}^{\:(n+1)\perp}\right)\nonumber
\end{equation}
The preceding computation selects the set of unavailable DOA
estimates that have a higher likelihood at each iteration. Then, the
set of estimated DOAs corresponding to the optimum value of
\text{$\mu $} that minimizes \eqref{SML_objective_function} also at
each $n^{th}$ iteration is determined. Finally, the output of the
proposed MS-KAI-Nested-MUSIC algorithm is formed by the set of the estimates
obtained at the $I^{th}$ iteration, as described in Table
\ref{tab_MSKAI_Nes_MU}.
\begin{table}[htb!]
    \small
    \caption{MS-KAI-Nested-MUSIC algorithm}\smallskip
    \vspace{-1.0em}
    \scalebox{0.90}{
        \begin{tabular}{r l}
            \hline\\
            \multicolumn{2}{l}{\small $\textbf{\underline{Inputs}:}$}\\[0.7ex]
            \multicolumn{2}{l}{\small$\mathit{M_{1}}$,\hspace{2mm}$\mathit{M_{2}}$,\hspace{2mm}$\mathit{d}$,\hspace{2mm}$\lambda$,\hspace{2mm}$\mathit{N}$,\hspace{2mm}$\mathit{P}$ }\\[0.6ex]
            \multicolumn{2}{l}{\small\text{Received vectors}  $\bm {y}\left( 1\right) $,\hspace{2mm}$\bm{y} \left( 2\right) $,$\cdots$, $\bm {y}\left( N\right)  $}\\[0.6ex]

            \multicolumn{2}{l}{\small $\textbf{\underline{Outputs}:}$}\\[0.6ex]

            \multicolumn{2}{l}{\small\text{Estimates}\hspace{1mm}$\mathit{\hat{\theta}_{1}^{(n+1)}(\mu\,opt)}$,\hspace{2mm}$\mathit{\hat{\theta}_{2}^{(n+1)}(\mu\,opt)}$,$\cdots$,\hspace{2mm}$\mathit{\hat{\theta}_{P}^{(n+1)}(\mu\,opt)}$} \\[3.1ex]

            \multicolumn{2}{l}{\small $\textbf{\underline{First step}:}$}\\[0.9ex]

%

            \multicolumn{2}{l}{\small $\{\mathit{\hat{\theta}_{1}}^{(1)},\:\mathit{\hat{\theta}_{2}}^{(1)},\cdots,\mathit{\hat{\theta}_{P}}^{(1)}\}\;\;\underleftarrow{MUSIC}$ $\:(\bm{\widetilde{R}},P,d,\lambda)$}\\[1.2ex]

            \multicolumn{2}{l}{\small$\bm{\hat{A}}^{(1)}=\left[\bm{a}(\mathit{\hat{\theta}_{1}^{(1)}}),\bm{a}(\mathit{\hat{\theta}_{2}^{(1)}}),\cdots,\bm{a}(\mathit{\hat{\theta}_{P}^{(1)}})\right]$} \\ [1.1ex]

            \multicolumn{2}{l}{\small $\textbf{\underline{Second step}:}$}\\[0.9ex]

            \multicolumn{2}{l}{\small \textbf{for}  \textit{n}\hspace{1mm}=\hspace{2mm}\text{1}\hspace{1mm}:\hspace{1mm}\textit{I}}\\[0.6ex]

            \multicolumn{2}{l}{\small $\bm{\hat{Q}}_{A}^{(n)}= \bm{\hat{A}}^{(n)}\:(\bm{\hat{A}}^{(n)H}\:\bm{\hat{A}}^{(n)})^{-1}\:\bm{\hat{A}}^{(n)H}$ }\\[0.9ex]

            \multicolumn{2}{l}{\small$\bm{\hat{Q}}_{A}^{(n)\perp}=\bm{I}_{M}\:-\:\bm{\hat{Q}}_{A}^{(n)}$ }\\[1.0ex]

            \multicolumn{2}{l}{\small $\bm{V}^{(n)}=\bm{\hat{Q}}_{A}^{(n)}\:\bm{\widetilde{R}}\:\bm{\hat{Q}}_{A}^{(n)\perp}$}\\[1.8ex]

            \multicolumn{2}{l}{\small \textbf{for} $\bm{\mu}=\mathrm{\:0:\iota\::1}$}\\[0.9ex]

            \multicolumn{2}{l}{\small $ \bm{\widetilde{R}}^{(n+1)} = \bm{\widetilde{R}}\:-\:\mathrm{\mu}\:(\bm{V}^{(n)}\:+\:\bm{V}^{(n)H})$} \\[0.9ex]

            \multicolumn{2}{l}{\small $\{\mathit{\hat{\theta}_{1}}^{(n+1)},\:\mathit{\hat{\theta}_{2}}^{(n+1)},\cdots,\mathit{\hat{\theta}_{P}}^{(n+1)}\}\;\;\underleftarrow{MUSIC}$ $\:(\bm{\widetilde{R}}^{(n+1)},\:P,d,\lambda)$}\\[1.4ex]

            \multicolumn{2}{l}{\small$\bm{\hat{B}}^{(n+1)}=\left[\bm{a}(\mathit{\hat{\theta}_{1}^{(n+1)}}),\bm{a}(\mathit{\hat{\theta}_{2}^{(n+1)}}),\cdots,\bm{a}(\mathit{\hat{\theta}_{P}^{(n+1)}})\right]$} \\ [1.4ex]

            \multicolumn{2}{l}{\small $\bm{\hat{Q}}_{B}^{(n+1)}= \bm{\hat{B}}^{(n+1)}\:(\bm{\hat{B}}^{(n+1)H}\:\bm{\hat{B}}^{(n+1)})^{-1}\:\bm{\hat{B}}^{(n+1)H}$}\\[1.2ex]

            \multicolumn{2}{l}{\small$\bm{\hat{Q}}_{B}^{(n+1)\perp}=\bm{I}_{M}\:-\:\bm{\hat{Q}}_{B}^{(n+1)}$}\\[1.2ex]

            \multicolumn{2}{l}{\small $\mathit{U^{(n+1)}(\mu)}=\mathrm{ln\:det}\left(\cdot\right),$}\\[1.1ex]

            \multicolumn{2}{l}{\small$\left(\cdot\right) =\left(\bm{\hat{Q}}_{B}^{(n+1)}\:\bm{\widetilde{R}}\:\bm{\hat{Q}}_{B}^{(n+1)}+\dfrac{{\rm Trace}\{\bm{\hat{Q}}_{B}^{\perp\:(n+1)}\:\bm{\widetilde{R}}\}} {\mathit{L-P}}\:\bm{\hat{Q}}_{B}^{\:(n+1)\perp}\right)$}\\[1.1ex]

            \multicolumn{2}{l}{\small $\mathit{\mu}_{\mathrm{o}}^{(n+1)}=\arg \min \hspace{1mm}\mathit{U^{(n+1)}(\mu)}$}\\[1.0ex]

            \multicolumn{2}{l}{\small $\mathrm{DOAs}^{(n+1)}= \left(\ast\right),$}\\[1.1ex]

            \multicolumn{2}{l}{\small$\left(\ast\right) =
                \{\mathit{\hat{\theta}_{1}^{(n+1)}(\mu_{o})}$,\hspace{2mm}$\mathit{\hat{\theta}_{2}^{(n+1)}(\mu_{o})}$,$\cdots$,\hspace{2mm}$\mathit{\hat{\theta}_{P}^{(n+1)}(\mu_{o})}\}$}\\[1.0ex]

            \multicolumn{2}{l}{\small$ \textbf{if}\hspace{1mm}  \mathit{n<=\;P}$}\\[0.6ex]

            \multicolumn{2}{l}{\small $\bm{\hat{A}}^{(n+1)}=\left\{\bm{a}(\mathit{\hat{\theta}_{\{1,\cdots,n\}}^{(n+1)}(\mu_{o})})\right\}\bigcup\left\{\bm{a}(\mathit{\hat{\theta}_{\{1,\cdots,P\}\,-\,\{1,\cdots,n\}}^{(1)}})\right\}$}\\[1.0ex]

            \multicolumn{2}{l}{\small \textbf{else}}\\[1.0ex]

            \multicolumn{2}{l}{\small$\bm{\hat{A}}^{(n+1)}= \left(\ast\right),$}\\[1.1ex]

            \multicolumn{2}{l}{\small$\left(\ast\right) =
                \left[\bm{a}(\mathit{\hat{\theta}_{1}^{(n+1)}}(\mu_{o})),\bm{a}(\mathit{\hat{\theta}_{2}^{(n+1)}}(\mu_{o})),\cdots,
                \bm{a}(\mathit{\hat{\theta}_{P}^{(n+1)}}(\mu_{o}))\right]$} \\ [1.4ex]

            \multicolumn{2}{l}{\small \textbf{end if}}\\[1.0ex]
            \multicolumn{2}{l}{\small \textbf{end for}}\\[1.0ex]

            \multicolumn{2}{l}{\small \textbf{end for}}\\[1.0ex]

            \hline
        \end{tabular}
    }
    \label{tab_MSKAI_Nes_MU}

\end{table}
\section{Computational Complexity Analysis}
\label{computational_analysis}

    In this section, we evaluate the approximate computational cost of the proposed
    MS-KAI-Nested-MUSIC algorithm in terms of multiplications and additions. For this purpose, we make use of Table \ref{tab_MSKAI_Nes_MU_comp_burd}, where     $\mathrm{\tau}=\frac{1}{ \iota} +1$. The increment ${ \iota}$ is defined in Table \ref{tab_MSKAI_Nes_MU}.
 From Table \ref{tab_MSKAI_Nes_MU_comp_burd}, it can be seen that assuming the specific configuration used in the simulations \ref{simulations}, MS-KAI-Nested-MUSIC  shows  a roughly similar computational burden in terms of multiplications and also of additions with

    $\mathcal{O} \left\lbrace I\tau \left[ \frac{180}{\Delta} \left( \frac{M^{2}}{4} +\frac{M}{2} \right)^{2}          \right] + \left( \frac{M^{2}}{4} +\frac{M}{2} \right) 8N^{2}                    \right\rbrace  $,
    where $\tau$
    is typically an integer that ranges from $ 1$  to $20 $, $\Delta$  stands for the search step and \textit{$ I $} is the number of iterations  at the $ 2nd $ step. The relatively high costs come from the two nested loops for
    computing $I\times\tau$ times two subprocesses at its second step.  These nested loops, from which the last is  the more significant, concentrate most of the required  operations. For this reason it is responsible for most of the burden of the proposed MS-KAI-Nested-MUSIC algorithm.

\begin{table}[htb!]
    \small
    \caption{MS-KAI-Nested-MUSIC algorithm}\smallskip
    \vspace{-1.0em}
    \scalebox{0.90}{
        \begin{tabular}{ l}
        \hline\\
        \underline{Multiplications} \\[2pt]
        \rule{0pt}{3ex}
        $\approx\: I\tau \left\lbrace \frac{180}{\Delta}[\left( \frac{M^{2}}{4} +\frac{M}{2} \right)^{2} +\left( \frac{M^{2}}{4} +\frac{M}{2} \right)\left( 2-P\right) -P]\right. $\\[1pt]

        $\left. + \left( \frac{M^{2}}{4} +\frac{M}{2} \right)8N^{2}
        {+\frac{10}{3}\left( \frac{M^{2}}{4} +\frac{M}{2} \right)^{3}}  +\left( \frac{M^{2}}{4} +\frac{M}{2} \right)^{2} \left(P+2 \right) \right.$\\$\left.+ \left( \frac{M^{2}}{4} +\frac{M}{2} \right)\left(P^{2}+2P \right)
          +\frac{P^{3}}{2}+\frac{3P^{2}}{2}  \right\rbrace $\\[1pt]

         \underline{Additions} \\[2pt]

        \rule{0pt}{3ex}
        $\approx\:I\tau \left\lbrace \frac{180}{\Delta}[\left( \frac{M^{2}}{4} +\frac{M}{2} \right)^{2} -\left( \frac{M^{2}}{4} +\frac{M}{2} \right)\left( P-1\right) ]\right. $\\[1pt]

        $\left. + \left( \frac{M^{2}}{4} +\frac{M}{2} \right)8N^{2}
        {+\frac{10}{3}\left( \frac{M^{2}}{4} +\frac{M}{2} \right)^{3}}  +\left( \frac{M^{2}}{4} +\frac{M}{2} \right)^{2} \left(P-1 \right) \right.$\\$\left.+ \left( \frac{M^{2}}{4} +\frac{M}{2} \right)\left(\frac{3P^{2}}{2}+   \frac{5P}{2} -1 \right)
        -P^{2}-\frac{P}{2}  \right\rbrace $\\[1pt]
        \hline

        \end{tabular}
}
\label{tab_MSKAI_Nes_MU_comp_burd}
\end{table}

%
%
%
%
%
%
%
%
%

\section{Simulations}
\label{simulations}

In this section, we examine the performance  of the proposed
MS-KAI-Nested-MUSIC algorithm in terms of probability of resolution (PR) and
RMSE  and compare them to the corresponding performances of Nested-MUSIC \cite{Pal1} and  of the original MUSIC \cite{Schimdt}. We focus on the specific case of closely-spaced sources.

 We employ $ M=8 $ sensors in the algorithms based on  two-level nested array and, in the original MUSIC, we use a ULA with  $ M_{1}=20$ sensors, which is also the  same number of sensors $\left(  M^{2}/4+ M/2 \right) $  of the filled  ULA obtained from  part of  the difference coarray, which is the actual number of sensors employed the Nested-MUSIC and MS-KAI-Nested-MUSIC algorithms.
 We assume an  inter-element
spacing $\Delta=\frac{\lambda_{c}}{2}$ and also that  there are two
uncorrelated complex Gaussian signals with equal power impinging on
the arrays. The closely-spaced sources are separated by
\textit{$2^{o}$}, at $\mathrm (15^{o},17^{o})$. The first two figures make use of $ N=150 $ snapshots and $L_{r}=250$ trials, whereas the two later ones  employ $ 3.33 dB $ and $L_{r}=250$ trials.

\begin{figure}[!h]

    \centering 
    \includegraphics[width=8.4cm, height=6.3cm]{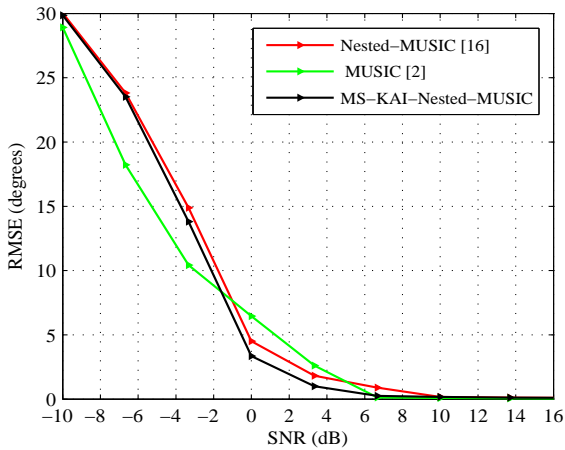} 
    \vspace{-1.0em}\caption{RMSE in degrees versus SNR with with $P=2$, $M=8$, $N=150$, $L_{r}=250$ runs.}
    \label{figura:RMSE_nest_unc_2sour_2deg_8sens_150snap_250runs}
\end{figure}

\begin{figure}[!h]

    \centering 
    \includegraphics[width=8.4cm, height=6.3cm]{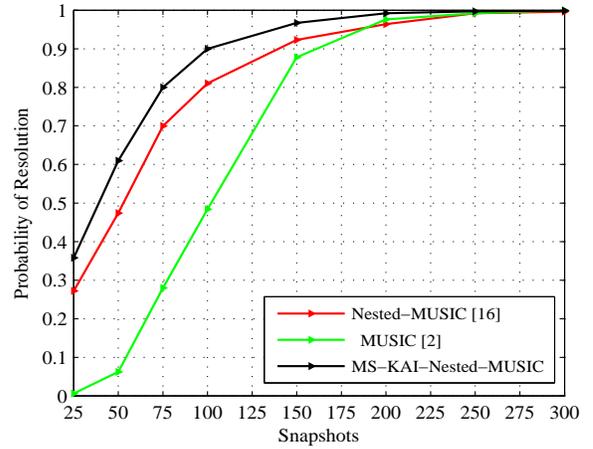} 
    \vspace{-1.0em}\caption{Probability of resolution versus SNR with $P=2$, $M=8$, $SNR=3.33\,dB$, $L_{r}=500$ runs.}
    \label{figura:PR_nest_unc_2sour_2deg_8sens_3pto3dB_500runs}
\end{figure}

In Fig.
\ref{figura:PR_nest_unc_2sour_2deg_8sens_3pto3dB_500runs}, we
show the probability of resolution versus SNR. We take into account
the criterion \cite{Stoica3}, in which two sources with DOAs
$\theta_{1}$ and $\theta_{2}$  are said to be resolved if their
respective estimates $\hat{\theta}_{1}$ and $\hat{\theta}_{2}$ are
such that both $\left|\hat{\theta}_{1} -\theta_{1}\right|$ and
$\left|\hat{\theta}_{2} -\theta_{2}\right|$ are less than
$\left|\theta_{1} -\theta_{2}\right|/2$. It can be seen  the
superior performance of the proposed MS-KAI-Nested-MUSIC in the range $\left(-10\;7 \right) dB $. From this point on, all considered algorithms
provide similar performance. The gap between the proposed MS-KAI-Nested-MUSIC and the Nested-MUSIC \cite{Pal1} shows a significant improvement achieved in terms of PR. It can be noticed a bigger gap between the proposed MS-KAI-Nested-MUSIC and the original MUSIC \cite{Schimdt}, whose number of physical sensors is $ 2.5\times $ the number of the physical sensors  of the other two-level nested based algorithms under comparison, what means an important saving of sensors.
\begin{figure}[!h]

    \centering 
    \includegraphics[width=8.4cm,height=6.3cm]{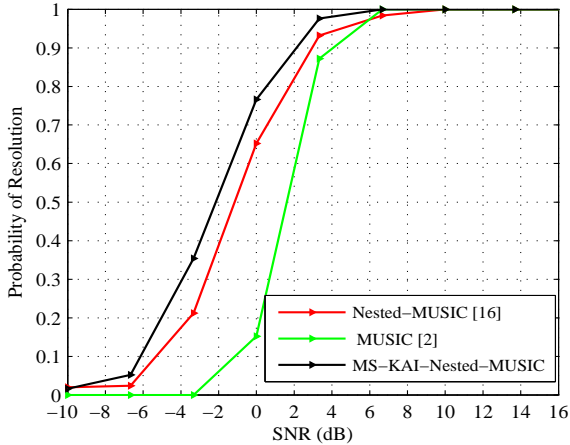} 
    \vspace{-1.0em}\caption{Probability of resolution versus SNR with $P=2$, $M=8$, $N=150$, $L_{r}=250$ runs}
    \label{figura:PR_nest_unc_2sour_2deg_8sens_150snap_250runs}
\end{figure}
In Fig. \ref{figura:RMSE_nest_unc_2sour_2deg_8sens_150snap_250runs},
it is shown the RMSE in degrees versus SNR. The RMSE is defined as
\begin{equation}
\centering \mathrm{RMSE}
=\sqrt{\frac{1}{L_{r}\:P}\sum\limits^{L_{r}}_{l=1}\sum\limits^{P}_{p=1}\bm(\theta_p
-\bm \hat{\theta}_p(l))^{2}}, \label{RMSE_run}
\end{equation}
where $ L_{r} $ is the number of trials.

It can be noticed that the MS-KAI-Nested-MUSIC outperforms Nested-MUSIC,
in the whole range under consideration.
In the range $\left[-10\;-1.8 \right) dB $., it is outperformed by  conventional  MUSIC, however, the achieved  levels of RMSE are still.  From $ -1.8 $ to $ 6.7\,dB $ MS-KAI-Nested-MUSIC is superior to it. From $ 10\,dB  $ on all algorithms have similar performance. As mentioned  before, it must be highlighted that in this specific case MUSIC makes use of a ULA whose number of physical sensors is  $ 2.5\times $ the number of the physical sensors  of the other two-level nested based algorithms under comparison.

\begin{figure}[!h]

    \centering 
    \includegraphics[width=8.4cm, height=6.3cm]{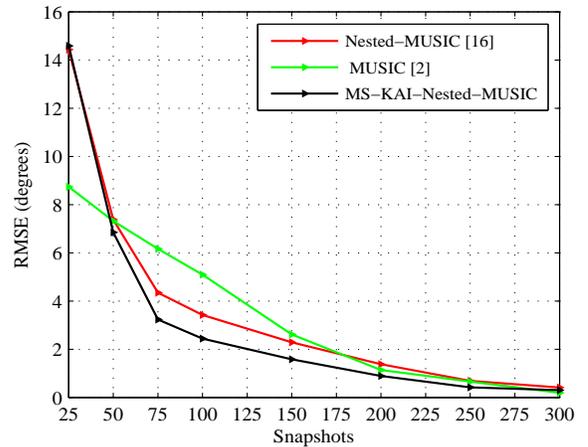} 
    \vspace{-1.0em}\caption{RMSE versus SNR with $P=2$, $M=8$, $SNR=3.33\,dB$, $L_{r}=500$ runs.}
    \label{figura:RMSE_nest_unc_2sour_2deg_8sens_3pto3dB_500runs}
\end{figure}

In Fig.
\ref{figura:PR_nest_unc_2sour_2deg_8sens_3pto3dB_500runs},
it is shown the influence of the number of snapshots on the probability of resolution. For this purpose we have set the SNR at $3.33\,dB$ and employed $500$ trials.
From the curves, it can be noticed the superior performance in the range $\left[ 25\; 250\right) $ snapshots. From this upper bound on, all algorithms have the same performance.
In Fig.
\ref{figura:RMSE_nest_unc_2sour_2deg_8sens_3pto3dB_500runs},
it is shown the influence of the number of snapshots on RMSE. In this case, we also set the SNR at $3.33\,dB$ and employed $500$ trials. It can be seen  that the performance of the MS-KAI-Nested-MUSIC is superior to the Nested-MUSIC. It can also be noticed that except for the range $\left[ 25\; 50\right)$, in which the RMSE has high levels,  the performance of MS-KAI-Nested-MUSIC is also superior to the  the original MUSIC \cite{Schimdt}, whose number of physical sensors is $ 2.5\times $ the number of the physical sensors  of the other two-level nested based algorithms under comparison.

\section{Conclusions}
\label{conclusions}

We have proposed the MS-KAI-Nested-MUSIC algorithm which gradually exploits
the knowledge of source signals obtained on line and the structure
of the covariance matrix and its perturbations. MS-KAI-Nested-MUSIC
algorithm can obtain significant gains in RMSE or probability of
resolution performance over the original Nested-MUSIC, and has
excellent potential for applications with sufficiently large data records in large-scale antenna systems for wireless communications, radar and
other large sensor arrays.

\end{document}